# Room temperature charge-to-spin conversion from q-2DEG at SrTiO$_3$-based interfaces

*Utkarsh Shashank[1,#], Angshuman Deka[1,2,#], Chen Ye[3], Surbhi Gupta[4], Rohit Medwal[5], Rajdeep Singh Rawat[4], Hironori Asada[6], X. Renshaw Wang[3,7]\*, and Yasuhiro Fukuma[1,8]\*\**

[#]Equal contribution
\* E-mail: renshaw@ntu.edu.sg
\*\*E-mail: fukuma@phys.kyutech.ac.jp

Mr. Utkarsh Shashank, Dr. Angshuman Deka, Prof. Yasuhiro Fukuma
[1]Department of Physics and Information Technology, Faculty of Computer Science and System Engineering, Kyushu Institute of Technology, 680-4 Kawazu, Iizuka 820-8502, Japan.

Dr. Angshuman Deka
[2]Birck Nanotechnology Center, School of Electrical and Computer Engineering, Purdue University, West Lafayette, Indiana 47907, USA

Prof. Y. Fukuma
[8]Research Center for Neuromorphic AI hardware, Kyushu Institute of Technology, Kitakyushu 808-0196, Japan

Dr. Chen Ye
[3]School of Physical and Mathematical Sciences, Nanyang Technological University 637616, Singapore.

Prof. X. Renshaw Wang
[3]School of Physical and Mathematical Sciences, Nanyang Technological University 637616, Singapore.
[7]School of Electrical and Electronic Engineering, Nanyang Technological University, 639798, Singapore

Dr. Surbhi Gupta, Prof. Rajdeep Singh Rawat
[4]Natural Science and Science Education, National Institute of Education, Nanyang Technological University 637616, Singapore.

Prof. Rohit Medwal,

[5]Department of Physics, Indian Institute of Technology Kanpur, Kanpur 208016, India.

Prof. Hironori Asada,

[6]Graduate School of Sciences and Technology for Innovation, Yamaguchi University, 2-16-1 Tokiwadai, Ube 755-8611, Japan

**Abstract**

Interfacial two-dimensional electron gas (2DEG), especially the $SrTiO_3$-based ones at the unexpected interface of insulators, have emerged to be a promising candidate for efficient charge-spin current interconversion. In this article, to gain insight into the mechanism of the charge-spin current interconversion at the oxide-based 2DEG, we focused on conducting interfaces between insulating $SrTiO_3$ and two types of aluminium-based amorphous insulators, namely $SrTiO_3$/AlN and $SrTiO_3$/$Al_2O_3$, and estimated their charge-spin conversion efficiency, $\theta_{cs}$. The two types of amorphous insulators were selected to explicitly probe the overlooked contribution of oxygen vacancy to the $\theta_{cs}$. We proposed a mechanism to explain results of spin-torque ferromagnetic resonance (ST-FMR) measurements and developed an analysis protocol to reliably estimate the $\theta_{cs}$ of the oxide based 2DEG. The resultant $\theta_{cs}/t$, where $t$ is the thickness of the 2DEG, were estimated to be 0.244 nm$^{-1}$ and 0.101 nm$^{-1}$ for the $SrTiO_3$/AlN and $SrTiO_3$/$Al_2O_3$, respectively, and they are strikingly comparable to their crystalline counterparts. Furthermore, we also observe a large direct current modulation of resonance linewidth in $SrTiO_3$/AlN samples, confirming its high $\theta_{cs}$ and attesting an oxygen-vacancy-enabled charge-spin conversion. Our findings emphasize the defects' contribution to the charge-spin interconversion, especially in the oxide-based low dimensional systems, and provide a way to create and enhance charge-spin interconversion via defect engineering.

# 1. Introduction

The formation of 2-dimensional electron gas (2DEG) at oxide interfaces has attracted significant research interest owing to the presence of conduction electrons at the interfaces of two insulators and its interesting phenomena, such as superconductivity and magnetism [1-3]. In addition to the observation of interfacial conductivity, the ability to achieve high carrier mobility > $10^5$ cm V$^{-1}$s$^{-1}$ in 2DEG is particularly interesting in the development of all-oxide devices [4-5]. Theoretical reports [6-8] predicted that, due to the presence of an inversion asymmetry at the interface, a 2DEG electron with momentum $\vec{p}$ in the presence of an electric field $\vec{E}$ can experience a Rashba-like field $\propto \vec{p} \times \vec{E}$ in its rest frame because of spin-orbit coupling (SOC)[9]. The possibility of achieving a strong Rashba SOC at these 2DEG allows compatibility with spintronic devices, and, therefore, has given rise to a series of experiments to understand magnetotransport in the 2DEG [10-20]. A large number of these investigations have been focused on spin-charge interconversion in 2DEG formed at interfaces of epitaxial LaAlO$_3$ (LAO) and LaTiO$_3$ (LTO) films grown on single-crystalline SrTiO$_3$ (STO) substrates [15-20]. Recently, reports on giant room temperature charge to spin conversion efficiency, $\theta_{cs}$ as high as 6.3 in epitaxial LAO/STO-based 2DEG [15], implying that they are significantly more efficient than most heavy metals (HM) [21], topological insulators [22-23] and engineered HM [24-26] towards spin current generation. Such properties position the 2DEG as an interesting candidate for applications in future spintronic devices, such as power-efficient spin-charge current interconversion.

Recent reports on high mobility in these 2DEG/amorphous oxides formed in systems like STO/Al$_2$O$_3$ interfaces have brought back attention towards 2DEG formed in such materials owing to their relative ease of deposition compared to their crystalline counterparts [4,27]. These 2DEG originate from oxygen vacancies in the STO side and are confined to a few nanometers (nm) inside the substrate surface. Broadly speaking, when a metal-based oxide/nitride, especially the Al-based materials, is deposited on an STO substrate, a layer of oxygen vacancies is created at the surface of STO due to the redox reactions between Al-based oxide/nitride and STO [28-30]. Such oxygen vacancies lead to mobile electrons, producing interfacial quasi-2DEG (hereafter, q-2DEG) inside STO. This leads to the creation of a conductive channel inside the STO substrate, as schematically shown in Fig. 1(a). Such q-2DEG induced by oxygen vacancy have often been overlooked in earlier studies and therefore we focus on potential higher charge-spin interconversion in these systems. Furthermore, although spin-charge current interconversion in 2DEG was reported at cryogenic temperatures[27], a systematic analysis of such a phenomenon at room temperature is lacking. Previous studies have indicated that angular-dependent measurement of magnetization dynamics is crucial towards an accurate estimation of magnetic anisotropies and torques [26, 31-34]. Therefore, we use such an approach to systematically investigate the charge-to-spin conversion efficiency of q-2DEG.

In this study, using spin torque ferromagnetic resonance (ST-FMR) measurements in STO/AlN/NiFe and STO/Al$_2$O$_3$/NiFe systems, we show that the $\theta_{cs}$ in the q-2DEG can be as high as their crystalline counterparts at room temperature. Our experiments clearly show the presence of a symmetric component in the ST-FMR lineshape, which has an odd parity with the direction of the magnetic field, confirming the presence of spin-torque in the devices. In addition to the expected $\sin 2\phi \cos\phi$ dependence of the symmetric and antisymmetric amplitudes of the ST-FMR spectra, we also detect contributions from a $\sin 2\phi \sin\phi$-like and $\sin 2\phi$-like behaviour. By filtering out the $\sin 2\phi \cos\phi$ contributions of the ST-FMR spectra, we estimate $\theta_{cs}/t = 0.244$ nm$^{-1}$ and $0.101$ nm$^{-1}$ for the STO/AlN/NiFe and STO/Al$_2$O$_3$/NiFe systems, respectively. Additionally, using DC-biased ST-FMR measurements, we show that modulation of resonant linewidth in the STO/AlN/NiFe samples is ~3 times higher than conventional heavy metals, *i.e.* Pt, directly affirming the presence of a high in-plane damping-like spin torque arising from the q-2DEG.

## 2. Experimental details

A pattern of dimension 80×20 μm² was created on TiO$_2$-terminated STO (100) substrates using photolithography, wherein 10 nm AlN or Al$_2$O$_3$ was deposited using pulsed laser deposition (PLD) at room temperature at a pressure of 10$^{-4}$ Pa. The sheet resistance, $R_s$ of the STO/AlN and STO/Al$_2$O$_3$ samples are ~90 kΩ/□. and ~10$^3$ kΩ/□, respectively. Subsequently, 5 nm NiFe was deposited using DC sputtering on the same pattern. Finally, using aligned photolithography and DC sputtering, Ti (10 nm)/Al (200 nm) electrodes were deposited on either side of the 80×20 μm² pattern to create the device, as shown in Fig. 1(b). See the supplementary information S1 for details of device fabrication. ST-FMR spectra in the devices were recorded using a lockin-detection technique (see Fig. 1(b)). A microwave current $I_{rf}$ of frequency $f$ was applied in the device using a signal generator while an external magnetic field $H_{ext}$ was swept in the in-plane direction. The $H_{ext}$ could be rotated in-plane with respect to the $I_{rf}$, as depicted in the right panel of Fig. 1(a) by varying the angle $\phi$ with the $I_{rf}$. All measurements were performed at room temperature.

## 3. Results and discussion

Due to the SOC at the q-2DEG, a spin current is generated in q-2DEG, which flows to the adjacent NiFe layer and exerts a torque on its magnetization. This can lead to FMR in the NiFe layer. A combined effect of an alternating anisotropic magnetoresistance (AMR) of NiFe and the microwave current $I_{rf}$ (proportional to spin current), gives the rectified DC voltage $V_{mix}$, which is recorded using a lock-in amplifier (Fig. 1(b)). A typical spectrum obtained from this measurement at a frequency of 4.0 GHz is shown in Fig. 1(c). This spectrum is then fitted with the sum of a symmetric Lorentzian and an antisymmetric component using the following equation [35].

$$V_{mix} = SF_{sym}(H_{ext}) + AF_{asym}(H_{ext}), \qquad (1)$$

where $F_{sym}(H_{ext}) = \frac{(\Delta H)^2}{(H_{ext}-H_o)^2+(\Delta H)^2}$, is the symmetric component with weight $S$, $F_{asym}(H_{ext}) = \frac{\Delta H(H_{ext}-H_o)}{(H_{ext}-H_o)^2+(\Delta H)^2}$, is the antisymmetric component with weight A, and ΔH and $H_0$ are the half-width-at-half-maximum and resonance field of the FMR spectra [25,35]. From the fit, we obtain $\mu_o\Delta H$= 4.34 mT for STO/AlN/NiFe and $\mu_o\Delta H$= 1.71 mT for the STO/Al$_2$O$_3$/NiFe sample at $f$ = 5 GHz, which indicates lower damping for the latter sample. (also see supplementary information S2). Meanwhile, we obtained $\mu_oH_0$= 38.54 mT for STO/AlN/NiFe and $\mu_oH_0$= 33.59 mT for the STO/Al$_2$O$_3$/NiFe sample at f= 5 GHz. Because the NiFe layers were deposited on both samples simultaneously, the difference in resonance field may be due to different effective demagnetizing fields, $M_{eff}$, in the two samples (Also see the Supplementary information S2) [33].

In addition, we also observed that, while the symmetric component magnitude is almost similar with an opposite sign with the $H_{ext}$ reversed, the antisymmetric component is different upon $H_{ext}$ reversal. The odd parity of the symmetric component with respect to the direction of the magnetic field indicates that it originates from the spin-torque generated by the q-2DEG. Due to the SOC at the q-2DEG, a spin current is generated transverse to the flow of microwave current $I_{rf}$. This spin current flows from the q-2DEG to the NiFe layer and exerts a torque on the magnetization of NiFe, resulting in the symmetric component of the ST-FMR spectra. Additionally, the $I_{rf}$ also generates an Oersted field, $h_{rf}$ which can exert a field-torque on the magnetization of NiFe layer and give rise to the antisymmetric component of the ST-FMR spectra [25]. In order to avoid any contributions to the rectified voltage lineshape that can arise from non-linear excitation of magnetization dynamics, we ensured that all our measurements were performed in the linear regime by checking the input power $P_{app}$ dependence of the spectra [33, 34]. As power increases from 0 to 13 dBm, $\Delta H$ and $H_0$ remains invariant with power, as shown in Supplementary information S3. This confirms that the $V_{mix}$ is due to the excitation of uniform

FMR mode in this input power range. We performed our measurements at $P_{app}$ = 10 dBm (10 mW) in order to ensure that the measurements are in the linear regime.

Frequency-dependent measurements of the ST-FMR spectra in the range of 3.5 - 5 GHz for the STO/AlN/NiFe and STO/Al$_2$O$_3$/NiFe samples were performed (see Fig. 2). Throughout the frequency range, we observed a similar behaviour of the symmetric and antisymmetric components of the ST-FMR spectra for both samples as described in the previous paragraph. The amplitude of the spectra decreases at a higher frequency. The resonance field, $H_o$ increases increasing frequency, which in turn leads to a lower precession cone angle at higher $H_{ext}$. Additionally, the $H_o$ increases with the frequency, which agrees well with the Kittel equation. Upon fitting to the Kittel equation [34], we obtained $\mu_o M_{eff}$ = 698 mT and 814 mT for the STO/AlN/NiFe and the STO/Al$_2$O$_3$/NiFe samples, respectively (see Supplementary information S2). This confirms that the difference in resonance fields in STO/AlN/NiFe and STO/Al$_2$O$_3$/NiFe samples arises from the different effective demagnetizing fields [34].

When the symmetric component of the ST-FMR spectra is attributed to a y-polarized spin current, $\hat{\sigma}_y$ travelling in the z-direction, the angular dependence of the symmetric amplitude follows a $\sin2\phi\cos\phi$ dependence. Simultaneously, due to the flow of an x-axial microwave current in the device, an antisymmetric component is also produced by the corresponding Oersted field, which follows a $\sin2\phi\cos\phi$ dependence. In this scenario, the $\theta_{cs}$ can be estimated using [35]:

$$\theta_{cs} = \frac{S}{A}\frac{e\mu_o M_s dt}{\hbar}\sqrt{1 + \frac{M_{eff}}{H_o}}, \qquad (2)$$

where $t$ is the q-2DEG thickness, e is the elementary charge, $\mu_o$ is the permeability of free space, $M_S$ is the saturation magnetization of NiFe, $d$ is NiFe thickness, ℏ is the reduced Planck constant and $M_{eff}$ is effective magnetization obtained from the Kittel fitting. The corresponding angular dependent components of voltages in the STO/AlN/NiFe and STO/Al$_2$O$_3$/NiFe samples are plotted in Fig. 3(a) and 3(c), respectively. The symmetric component is fitted to the following equation (see Supplementary information S4 for details):

$$V_s = a\sin2\phi\cos\phi + b\sin\phi\sin2\phi, \quad (3)$$

where a is the weightage of the $\sin2\phi\cos\phi$ component and b is the weightage of the $\sin\phi\sin2\phi$ component. Meanwhile, the antisymmetric component is fitted to:

$$V_A = c\sin2\phi\cos\phi + d\sin\phi\sin2\phi + é\sin2\phi, \quad (4)$$

where $c$, $d$ and $é$ are the weights of the $\sin2\phi\cos\phi$, $\sin\phi\sin2\phi$ and $\sin2\phi$ components, respectively. We note that, in our devices, the angular dependence of symmetric and antisymmetric components is not purely $\sin2\phi\cos\phi$. Figures 3(b) and 3(d) show the weightage of the different symmetric and antisymmetric parts of the spectra for $f$ = 3.5-5 GHz in the case of the STO/AlN/NiFe and STO/Al$_2$O$_3$/NiFe samples, respectively. For STO/AlN/NiFe, the symmetric component has a 95.3% $\sin2\phi\cos\phi$ dependence, with the rest 4.7% arising from a $\sin2\phi\sin\phi$ dependence when averaged over the entire frequency range. Meanwhile, its antisymmetric component has an average of 49.5%, 22.5% and 28.0% contributions from a $\sin2\phi\cos\phi$, $\sin2\phi\sin\phi$ and $\sin2\phi$ dependence, respectively. In the case of the STO/Al$_2$O$_3$/NiFe, we find that the symmetric component has 75.5% $\sin2\phi\cos\phi$ dependence and 24.5% $\sin2\phi\sin\phi$ dependence averaged over the entire frequency range. On the other hand, its antisymmetric component has an average of 55.7%, 16.4% and 27.9% contributions from $\sin2\phi\cos\phi$, $\sin2\phi\sin\phi$ and $\sin2\phi$ dependences, respectively. This indicates breaking of the twofold (180°+$\phi$) and mirror (180°-$\phi$) symmetries of torques for both the q-2DEG-NiFe samples. Therefore the lineshape analysis method, *i.e.*, Eq. (2), which uses a spectrum obtained at a single azimuthal angle $\phi$, may not reveal the comprehensive picture of torques, leading to inaccurate quantification of SOTs [36].

Although an investigation into the exact origins of these additional components in the ST-FMR spectra is beyond the scope of this article, we would like to emphasize that we reproducibly observed this behaviour in multiple devices over a wide range of frequencies. This may be a consequence of non-uniform microwave current flow in devices. In order to rule out the possibility that it is caused by our device design, we further verified the same experiments in a Pt/NiFe device fabricated under similar conditions and found 100% sin2$\phi$cos$\phi$ dependence for both symmetric and antisymmetric components of the ST-FMR spectra (see Supplementary information S5). NiFe has a much lower resistivity compared to the q-2DEG created at the STO/AlN interface. This may lead to non-uniform current flow in the q-2DEG-NiFe device, giving rise to the additional angular-dependent components in the ST-FMR spectra for the STO/AlN/NiFe sample. It is noteworthy that the sin2$\phi$cos$\phi$ contribution reduces for both the symmetric and antisymmetric components in the case of STO/Al$_2$O$_3$/NiFe samples, whose 2DEG has ~10 times higher sheet resistance compared to that of STO/AlN/NiFe samples.

Before we proceed to the estimation of $\theta_{cs}$, it should be noted that $\theta_{cs}$ is directly proportional to the thickness of the q-2DEG. Therefore, without a direct measurement of the 2DEG thickness, we may end up with overestimated values of $\theta_{cs}$. In order to avoid such a discrepancy in our study, we estimate the q-2DEG thickness normalized conversion efficiency as seen from Eq. (5) below. Additionally, to estimate the $\theta_{cs}$ corresponding to the dominant y-polarized spin current, $\hat{\sigma}_y$ in the samples, we introduce a ratio a/c, where *a* and *c* are weight percentages of the sin2$\phi$cos$\phi$ component of the symmetric and antisymmetric parts at a specific frequency mentioned earlier. This allows us to rewrite Eq. (2) as follows:

$$\frac{\theta_{cs}}{t} = \left(\frac{a}{c}\right)\frac{S}{A}\frac{e\mu_o M_s d}{\hbar}\sqrt{1+\frac{M_{eff}}{H_o}}. \qquad (5)$$

Using Eq. (5), we estimated the median $\theta_{cs}/t$ to be 0.244 nm$^{-1}$ for the STO/AlN sample and 0.101 nm$^{-1}$ for the STO/Al$_2$O$_3$ sample. The frequency dependence of $\theta_{cs}/t$ is shown in Fig. 4. The values of $\theta_{cs}/t$ estimated using Eq. (2) are also plotted in the same figure to show the discrepancy between the two methods. Note that such a discrepancy is not seen for the Pt/NiFe devices (see Supplementary information S5).

As mentioned earlier, the calculated values in our case are normalized by the 2DEG thickness. Typically, in previous reports, the 2DEG thickness is assumed to be 10 nm, and a $\theta_{cs}$ ~ 1.8 at room temperature has been reported for STO/LaAlO3/NiFe [20], $\theta_{cs}$ ~ 2.4 for quasi-2DEG in an STO/LaTiO3/NiFe system [18] and $\theta_{cs}$ ~ 6.3 for STO/LaAlO3/CoFeB [15]. In our case, if we assume the 2DEG thickness, $t$ = 10 nm, the $\theta_{cs}$ comes out to be 2.44 for the STO/AlN/NiFe and 1.01 for STO/Al2O3/NiFe samples. These values are comparable to previous reports on 2DEG formed in epitaxial oxides grown on STO substrates. Note that the AlN and Al2O3 layers are amorphous in our case. Hence, the $\theta_{cs}$ created at the q-2DEG in our STO/amorphous oxide interfaces are shown to be comparable in terms of charge-to-spin conversion efficiencies with the 2DEG reported for the epitaxial oxides on STO. Moreover, the $\theta_{cs}$ in our sample is 1 or 2 orders of magnitude higher than that of HM, such as Pt [37], Ta [21], W [38], and also than engineered HM [24-26]. As seen from the sheet resistance values, the q-2DEG in STO/AlN has 1 order smaller resistance values compared to STO/Al2O3. This indicates that when AlN is deposited over STO substrates, higher number of $O^{2-}$ diffuse outward from the substrate to oxidize AlN compared to when Al2O3 is deposited. This in turn creates a larger number of oxygen defects inside the STO substrate. As demonstrated by Chen et al., this difference in diffusion may have to do with the difference in chemical reactivity of AlN and Al2O3 with TiO2-terminated STO[28]. Our observation of a higher charge-to-spin conversion in the STO/AlN samples indicates that the higher oxygen vacancies not only play a role in enhancing the electronic transport, but may also lead to a higher charge-to-spin conversion efficiency in the q-2DEG.

The modulation of damping of FMR can also be an additional tool to overcome the issues with $\theta_{cs}$ estimation mentioned in this paper as it provides a direct insight into the strength of damping-like spin torques. This is free from problems arising from impedance mismatch, unconventional spin current polarized in different directions, and Nernst heating [35]. In this method, an additional direct current (DC) $I_{dc}$ is applied along with $I_{rf}$. The spin current at the Rashba interface, which is proportional to $I_{dc}$, modulates the ferromagnetic resonance linewidth (or the Gilbert damping) [35] of NiFe via damping-like torque which has an odd polarity with $I_{dc}$. By measuring the change in linewidth as a function of $I_{dc}$, the $\theta_{cs}$ can be evaluated if the thickness of the q-2DEG is known. Therefore, we performed the $I_{dc}$-biased ST-FMR measurement on the STO/AlN/NiFe and STO/Al$_2$O$_3$/NiFe samples by applying an $I_{dc} \sim \pm$ 7mA along with $I_{rf}$. The data at $f$ = 5 GHz is shown in Fig. 5 (a) and 5(b). For the STO/AlN/NiFe, we observe a linear modulation of $\Delta H$ with $I_{dc}$. Upon changing the polarity of $I_{dc}$, we observe a reversed slope of $\Delta H$ vs $I_{dc}$ (see Fig. 5(a) and Supplementary information S6). This confirms the presence of dominant damping-like torque in STO/AlN/NiFe. The slope of linewidth modulation for STO/AlN/NiFe is ~3 times higher than the modulation seen for the Pt/NiFe samples as seen in Fig. 5(a) and 5(c). Due to a much lower resistivity of Pt, it implies that $\theta_{cs}$ for STO/AlN/NiFe will be ~1-2 orders higher compared to Pt/NiFe. However, we are unable to quantify $\theta_{cs}$ beyond its qualitative discussion since an accurate measure of q-2DEG resistivity is paramount for quantifying $\theta_{cs}$ from this measurement, and we do not know the exact thickness of q-2DEG. Furthermore, as seen in Fig. 5(b), we are unable to detect any linear modulation in linewidth for STO/Al$_2$O$_3$/NiFe. This is due to a significantly higher sheet resistance of the q-2DEG in STO/Al$_2$O$_3$, which results in very little current through its q-2DEG and difficulty in detecting any linear modulation of $\Delta H$ by $I_{dc}$. Instead, we obtain the dependence of $\Delta H$ vs $I_{dc}$ shown in Fig. 5(b), which can be attributed to the dominance of the heating effect over modulation of damping in the sample. An increase of $H_o$ with increasing $I_{dc}$ magnitude in STO/Al$_2$O$_3$/NiFe samples confirms the heating effect [22]. Notably, the $H_o$ does not change with $I_{dc}$ in either STO/AlN/NiFe or Pt/NiFe (see Supplementary information S7).

In summary, we report a large charge to spin conversion efficiency $\theta_{cs}$ at room temperature in 2DEG formed at the interface of amorphous AlN and Al$_2$O$_3$ with SrTiO$_3$ substrate. Previous reports on spin-charge interconversion have been reported for 2DEG formed in epitaxial thin films at room temperature [15-20] and amorphous oxide/STO substrates in 2DEG at cryogenic temperatures [27], which arises from the spin-momentum locking at Rashba interfaces lacking inversion symmetry. From angular-dependent ST-FMR measurements in our q-2DEG-NiFe samples, we observe the breaking of two-fold and mirror symmetry of spin-torques which may be due to non-uniform microwave current flow in the high impedance q-2DEG. Our estimation of $\theta_{cs}/t \sim 0.244$ nm$^{-1}$ in STO/AlN and $\sim 0.101$ nm$^{-1}$ in STO/Al$_2$O$_3$, comparable to that of epitaxial films is important because of the relative ease with which amorphous films can provide added functionalities in 2DEG applications [4,28-30]. Moreover, a large direct current modulation of linewidth for STO/AlN provides direct confirmation of high $\theta_{cs}$ in our q-2DEG. The similarity between our amorphous and crystalline oxide interfaces in terms of charge density and conductivity along with the charge-to-spin conversion may provide insight into the microscopic mechanisms and further optimization of the conversion efficiencies. An ability to efficiently generate spin current from q-2DEG, while simultaneously understanding their angular dependent properties that shed light on the directionality of spin torques, is crucial for their implementation in spintronic device applications.

**Supporting Information**

Supporting Information is available from the Wiley Online Library or from the author.

**Acknowledgements**

Y.F. and H.A. would like to acknowledge JSPS Grant-in-Aid (KAKENHI No. 22K04198), KIOXIA Corporation and Iketani Science and Technology Foundation. X.R.W. acknowledges support from Academic Research Fund Tier 2 (Grant No. MOE-T2EP50220-0005) and Tier 3 (Grant No. MOE2018-T3-1-002) from Singapore Ministry of Education.  R. M., S. G. and R. S. R. would like to acknowledge the support from the Ministry of Education, Singapore grant no MOE2019-T2-1-058 (ARC 1/19 RSR). R. M., S. G., R. S. R. and X.R.W. acknowledges support from the National Research Foundation (NRF), Singapore under its 21st Competitive Research Programs (CRP Grant No. NRF-CRP21-2018-0003).

**Figure Captions:**

**Figure** 1. (a) Schematic of q-2DEG formed inside the STO substrate at the interface with AlN or $Al_2O_3$. Upon deposition of AlN or $Al_2O_3$, the $O^{2-}$ diffuses outwards from the $TiO_2$-terminated STO, resulting in formation of a oxygen-vacancy induced quasi-2DEG (q-2DEG) near the surface of STO. The right panel additionally shows the multilayer structures and different parameters used in the measurement. (b) Schematic of ST-FMR set-up along with an optical image of the device. (c) ST-FMR spectra $V_{mix}$, together with fitted symmetric and antisymmetric components, which were obtained at f = 4 GHz for STO/AlN/NiFe and STO/$Al_2O_3$/NiFe.

**Figure** 2. Frequency-dependent ST-FMR spectra, for $f$ = 3.5-5 GHz, as a function of $H_{ext}$, for (a) STO/AlN/NiFe and (b) STO/$Al_2O_3$/NiFe.

**Figure 3.** Angular dependence of symmetric (Sym) and antisymmetric (Asym) components in ST-FMR spectra obtained as a function of $\phi$ for (a) STO/AlN/NiFe, and (c) STO/$Al_2O_3$/NiFe for $f$ = 4.0 GHz with solid lines (black and red) fit by Eq. (3) and Eq. (4), respectively. Weight (in %) of sin2$\phi$cos$\phi$ and sin2$\phi$sin$\phi$ for Symmetric (Sym) and sin2$\phi$cos$\phi$, sin2$\phi$sin$\phi$, sin2$\phi$ for Antisymmetric (Asym) components are obtained as a function of $f$ for (b) STO/AlN/NiFe and (d) STO/$Al_2O_3$/NiFe.

**Figure** 4. Thickness normalized charge to spin conversion efficiency $\theta_{cs}/t$ obtained from Eq. (2) and Eq. (5) as a function of $f$ for STO/AlN/NiFe and STO/$Al_2O_3$/NiFe.

**Figure** 5. DC-biased ST-FMR measurements for (a) STO/AlN/NiFe (b) STO/$Al_2O_3$/NiFe, and (c) Pt/NiFe showing $\mu_o \Delta H$ as a function of $I_{dc}$. The red and black lines depict the linear fitting in (a), (c) and non-linear curve in (b).

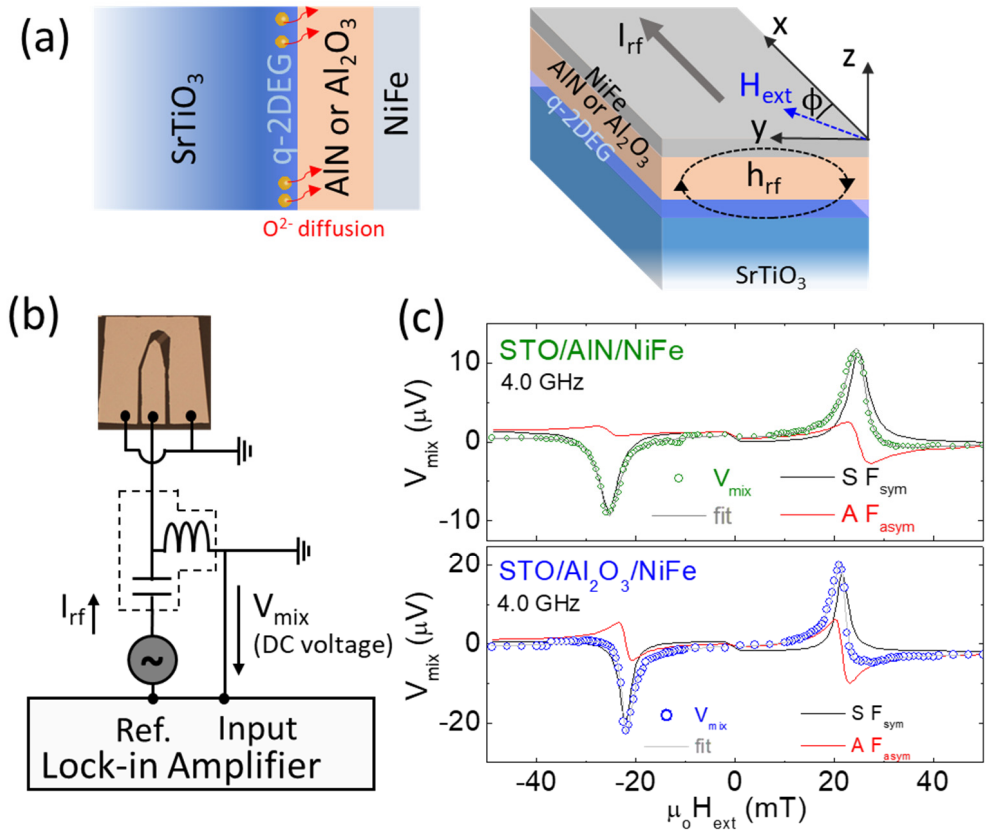

**Figure** 1. U. Shashank, A. Deka *et. al.*

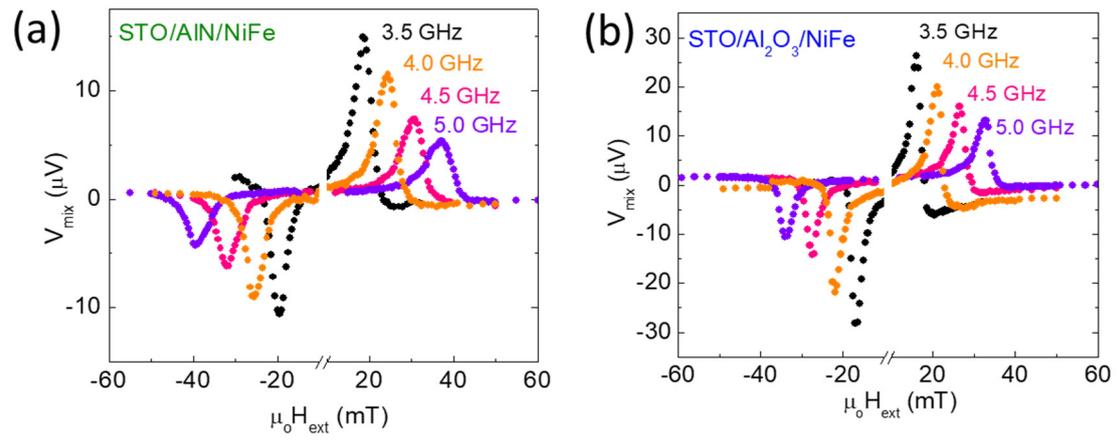

**Figure** 2. U. Shashank, A. Deka *et. al.*

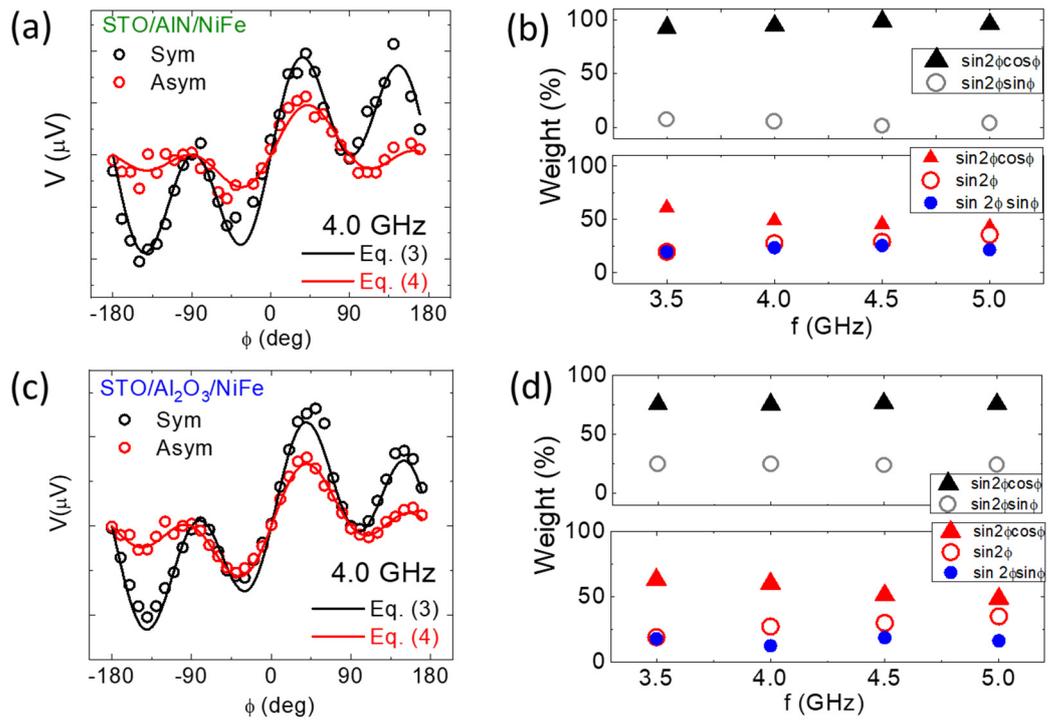

**Figure** 3. U. Shashank, A. Deka *et. al.*

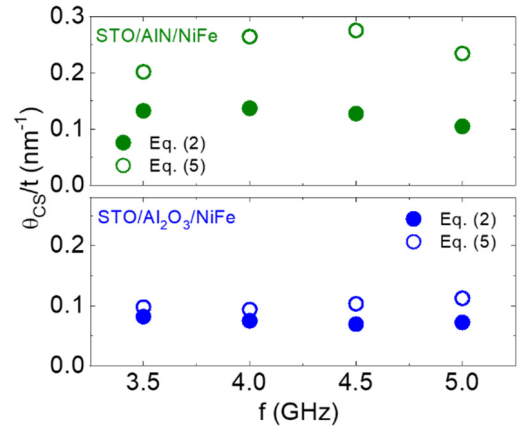

**Figure** 4. U. Shashank, A. Deka *et. al.*

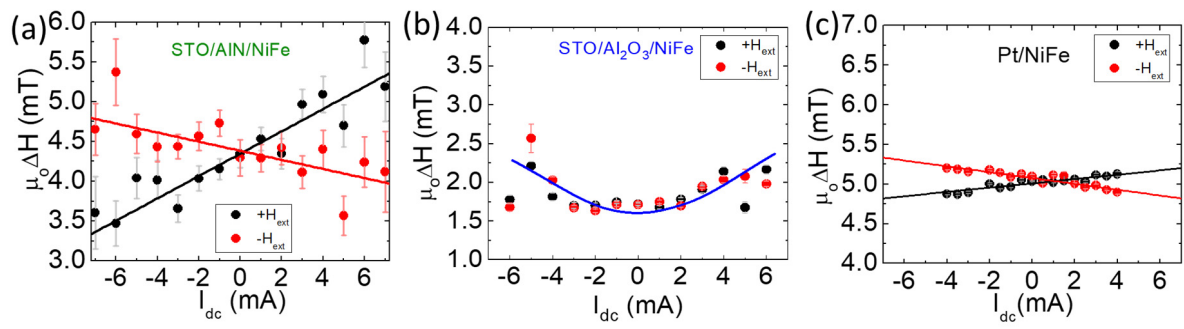

**Figure** 5. U. Shashank, A. Deka *et. al.*